\documentclass[12pt,preprint]{aastex}
\usepackage{emulateapj5}

\def\HI{H~{\sc i}} 

\def\kms{${\rm km~s^{-1}}$}

\shortauthors{MCCLURE-GRIFFITHS ET AL} 
\shorttitle{HI SHELLS BEHIND THE COALSACK}
\slugcomment{To appear in the Astrophysical Journal, vol. 562}

\begin{document} 

\title{\HI\ Shells Behind the Coalsack}
\author{N. M. McClure-Griffiths\altaffilmark{1} and John M. Dickey\altaffilmark{2}} 
\affil{Department of Astronomy, University of Minnesota, 116 Church Street
SE, Minneapolis, MN 55455}

\author{B. M. Gaensler\altaffilmark{3} }
\affil{Center for Space Research, Massachusetts Institute of Technology, 70
Vassar Street, Cambridge, MA 02139}

\and 

\author{A. J. Green\altaffilmark{4}} 
\affil{Astrophysics Department, School of Physics, University of Sydney, NSW
2006, Australia}

\altaffiltext{1}{\mbox naomi@astro.umn.edu}
\altaffiltext{2}{\mbox john@astro.umn.edu}
\altaffiltext{3}{Hubble Fellow; \mbox bmg@space.mit.edu}
\altaffiltext{4}{\mbox agreen@physics.usyd.edu.au}

\authoraddr{Address correspondence regarding this manuscript to: 
                N. M. McClure-Griffiths
                Department of Astronomy
                University of Minnesota
                116 Church St. S.E.
                Minneapolis, MN 55455}
\begin{abstract}
We report the discovery of two new large \HI\ shells in the direction of the
Coalsack nebula.  Both shells were observed with the Parkes Radio Telescope
as part of the Southern Galactic Plane Survey.  The largest shell,
GSH~304-00-12, is at a distance of $\sim 1.2$~kpc and has derived physical
dimensions of $280\times 200$~pc.  The second shell, GSH~305+01-24, is at a
distance of $\sim 2.2$~kpc and has derived dimensions of $280\times
440$~pc.  We present a simple numerical model to show that GSH~305+01-24
most likely formed from stellar winds in the Centaurus OB1 stellar
association.  There is associated radio, infrared and H$\alpha$ continuum
emission.  Both shells are situated in the Sagittarius-Carina arm, with
GSH~305+01-24 more distant.  The far edge of GSH~304-00-12 is at the near
side of the arm and opens into the interarm region.  We find no evidence for
closure at the near side of the shell and therefore describe the geometry as
conical.  Emission from the near side of the shell may be lost in absorption
by the Coalsack Nebula.
\end{abstract}

\section{Introduction}
\label{sec:intro}
\HI\ shells and supershells have long been studied in external galaxies, as
well as in our own Milky Way, as a probe of the star formation feedback
processes in the interstellar medium \citep[ISM; e.g.][]{westerlund66,
heiles79}.  These structures, which are often identified as voids in the
interstellar neutral hydrogen (\HI), range from tens to hundreds of parsecs
in diameter.  Most shells form from the combined effects of stellar winds
and supernovae that inject large quantities of energy ($10^{51} -
10^{53}$~ergs) into the ISM.  A few studies have directly identified the
shell power source as an OB association \citep[e.g.][]{saken92,brown95}.  In
these cases, however, standard wind blown bubble models
\citep[e.g.][]{weaver77} fail to properly account for the shells' radii and
expansion velocities.  As noted by \citet{oey95} the model typically
predicts radii that are too large and expansion velocities that are too
small when compared to the Large Magellanic Cloud (LMC) shell population.

In the nearby Large and Small Magellanic Clouds the large-scale structure of
the ISM is dominated by hundreds of expanding shells
\citep{staveley-smith97,kim98}.  Many of these have been studied in an
attempt to understand the dynamics of shells \citep{oey96a,oey96b,points99}
and their role in the ISM.  Unfortunately, much less is known about the
dynamics, density and distribution of \HI\ shells in the Milky Way.  Early
\HI\ surveys of the Milky Way resulted in the discovery of many shells
\citep{heiles79,heiles84}.  New high resolution surveys of the Galactic
Plane, such as the Canadian and Southern Galactic Plane Surveys
\citep{taylor99,mcgriff01a} are increasing this number and our knowledge of
the small-scale structure of Galactic \HI.  

We here report on the identification of two large shells, GSH~304-00-12 and
GSH~305+01-24, located at the edge of the nearby Sagittarius-Carina spiral
arm.  The Coalsack nebula lies between the Sun and the near edge of
GSH~304-00-12, though it does not appear physically related to the shell.
We believe that GSH~305+01-24 is associated with the Centaurus OB1 stellar
association.  We employ a very simple model for a wind-blown bubble to show
that the shell is well explained by the catalogued stars in Cen OB1.  In \S
\ref{sec:obs} we present the observations and data analysis.  In \S
\ref{sec:new shells} we describe the shells morphologically and dynamically
and suggest a geometry to explain their characteristics.  In \S
\ref{subsec:cenob1} we explore the relationship of GSH~305+01-24 with Cen
OB1.
\section{Observations and Analysis}
\label{sec:obs}
The data presented here were obtained as part of the Southern Galactic Plane
Survey \citep[SGPS;][]{mcgriff01a}.  These observations were made with the
multibeam system on the Parkes Radio Telescope, a 64~m antenna situated near
Parkes NSW, Australia\footnote{The Parkes telescope is part of the
Australia Telescope which is funded by the Commonwealth of Australia for
operation as a National Facility managed by CSIRO.}.  The Parkes portion
of the SGPS covers $253\arcdeg \leq l \leq 358\arcdeg$, $|b|=\pm 10\arcdeg$.
The observations were made during four observing sessions on 1998 December
15-16, 1999 June 18-21, 1999 September 18-27, and 2000 March 10-15.
Observations were made by the process of mapping ``on-the-fly'' with the
inner seven beams of the multibeam system.  The telescope was scanned
through three degrees in Galactic latitude while recording data in 5~s
samples.

These data were recorded in frequency switching mode using the narrow-band
back-end system \citep{haynes99}, with a total bandwidth at each frequency
of 8 MHz across 2048 channels.  The center frequency was switched between
1419~MHz and 1422.125~MHz every 5~s.  To remove the bandpass shape each
sample was divided by the previous frequency switched sample and the
residual bandpass shape fitted with a series of Fourier components.  The
spectra were then multiplied by the mean of the reference signal over the
spectrum to reconstruct the emission with a flat baseline.  This technique
for bandpass calibration has the advantage that continuum information is
retained but changes in the bandpass shape during the course of the
observing day are removed.  Absolute brightness temperature calibration of
the \HI\ line data was performed from daily observations of the IAU standard
regions S6 and S9 \citep{williams73}.  These data were shifted to the local
standard of rest (LSR) by applying a phase shift in the Fourier domain.  A
more detailed description of the Parkes observing strategy and calibration
is given by \citet{mcgriff00}.

The calibrated data were imaged using {\em Gridzilla}, a gridding tool
created for use with Parkes multibeam data and found in the ATNF package of
AIPS++\footnote{See http://aips2.nrao.edu/docs/aips++.html}.  The gridding
algorithm as implemented in {\em Gridzilla} is described by Barnes et al.\
(2001)\nocite{barnes01}.  The data presented here were gridded using a
weighted median technique, assuming a beamwidth of 16\arcmin.  A Gaussian
smoothing kernel of FWHM 18\arcmin\ was employed with a cutoff radius of
10\arcmin, and a cellsize of 4\arcmin.  The gridded data were imported to
the MIRIAD\footnote{R.\ J.\ Sault and N.\ E.\ B.\ Killeen, 2000, The Miriad
User's Guide (Sydney: Australia Telescope National Facility), at
http://www.atnf.csiro.au/computing/software/miriad/.}  data reduction
package for analysis \citep{sault00}.  Off-line channels in the ranges
$+160$ to $+200$~\kms\ and $-145$ to $-100$~\kms\ were used for continuum
subtraction with the MIRIAD task "IMLIN".  The final, calibrated data have
an angular resolution of 16\arcmin, a velocity resolution of $0.82$~\kms\
and a rms noise of $\sim 0.3$~K.

\section{\HI\ Shells GSH~304-00-12 and GSH~305+01-24}
\label{sec:new shells}
\begin{figure}
\begin{center}
\includegraphics[scale=0.6,angle=-90]{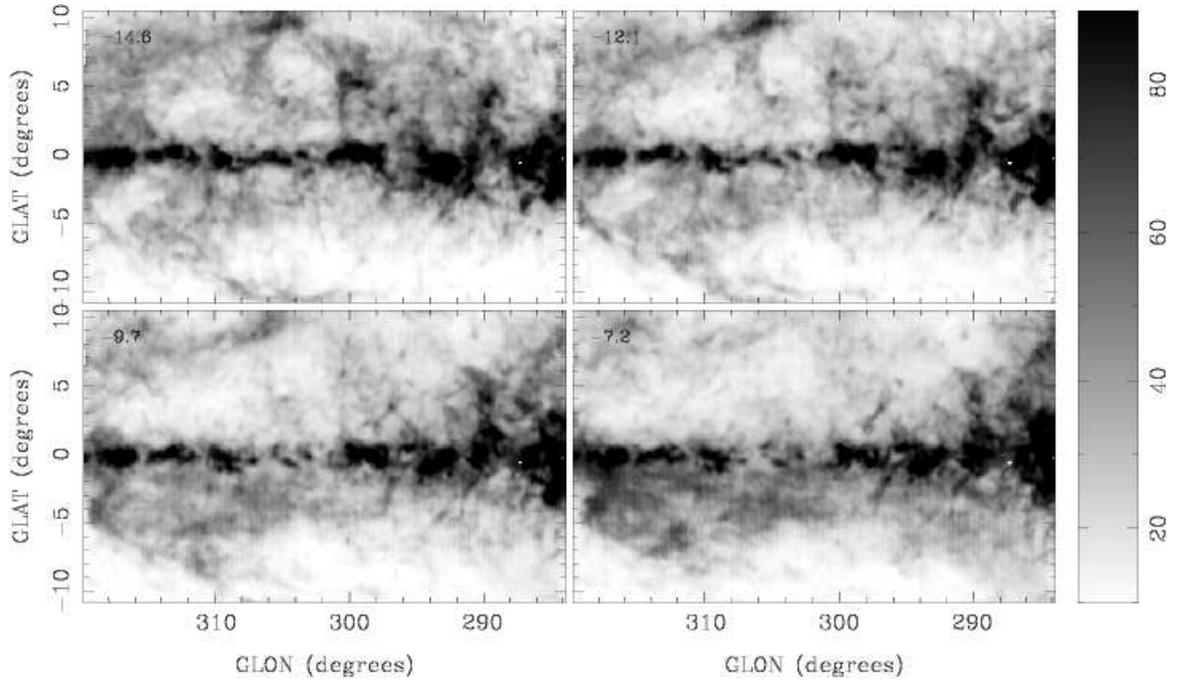} 
\figcaption[large]{Four velocity images of the largest shell, GSH~304+00-12,
  behind the Coalsack.  The images show velocity planes from $-14.8$~\kms\
  to $-7.2$~\kms\ and are separated by $2.5$~\kms.  The grey scale is linear
  from 10 - 90~K, as shown in the wedge at the right.  The shell, centered
  at $l=303\fdg9$ and $b=-0\fdg2$, extends to $b=\pm 10\fdg5$ at the center
  and is best seen as the arcs above and below the plane on the left of the
  images.
\label{fig:shell}}
\end{center}
\end{figure}

\begin{figure}
\begin{center}
\includegraphics[scale=0.8]{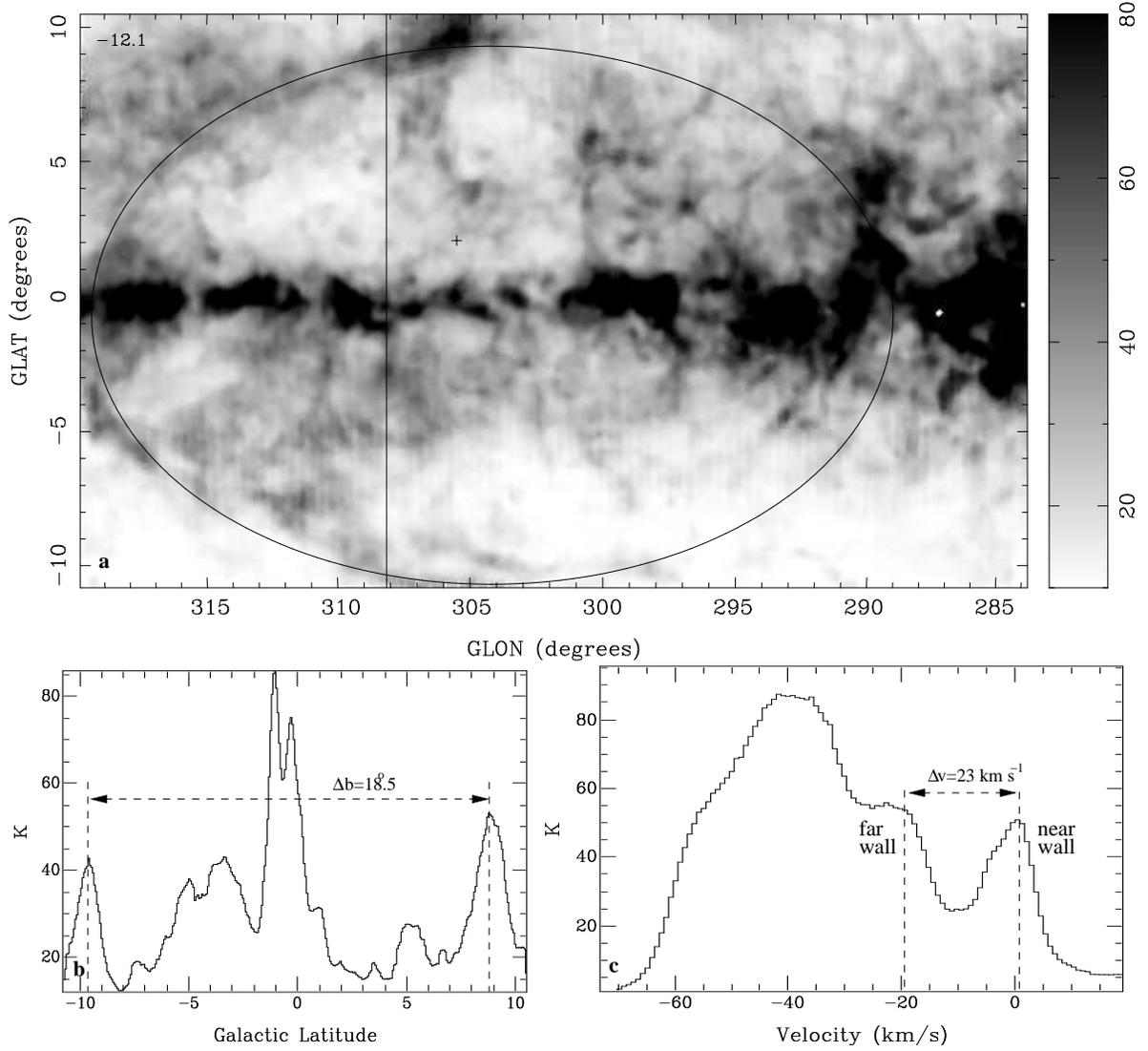} 
\figcaption[]{\HI\ channel image at $v=-12.1$~\kms\ of GSH~304+00-12 ({\bf
a}), latitude slice at $l=308\fdg15$ across the shell ({\bf b}), and a
velocity profile through the shell at $l=305\fdg43$, $b=2\fdg07$ ({\bf c}).
The outline of the shell is marked with the black ellipse on the \HI\
channel image.  The vertical line marks the position of the latitude cut
(panel {\bf b}) and the cross marks the position of the velocity profile
(panel {\bf c}).
\label{fig:annotated}}
\end{center}
\end{figure}

We here present images of a pair of new shells in the direction of the
Southern Coalsack.  The larger shell has an extremely large angular size of
$29\arcdeg \times 20\arcdeg$ and is centered at $l=303\fdg9$, $b=-0\fdg2$,
$v=-11.5$~\kms.  Following the IAU standard naming convention based on the
center coordinates of the shell, we name this shell GSH~304-00-12.  Velocity
channel images of GSH~304-00-12 from $-14.8$~\kms\ to $-7.2$~\kms\ in
intervals of $2.5$~\kms\ are shown in Fig.~\ref{fig:shell}.  The shell is a
large elliptical structure extending to $|b|\sim10\arcdeg$ in an arc that is
nearly symmetric about the plane.  For clarity, the shell is shown again in
Fig.~\ref{fig:annotated}, a composite images that includes an \HI\ image of
the shell at $v=-12$~\kms (panel a), a latitude cut across the shell (panel
b), and a velocity profile through the shell (panel c).  The \HI\ image is
marked with an ellipse that traces the edge of the shell.  There is a
vertical line marking the position of the latitude cut and a cross-hair
marking the position of the velocity profile.  

The shell first appears near $v\approx-27$~\kms, embedded in the \HI\
emission from the Sagittarius-Carina arm, where it is as an arc of length
$\sim 20\arcdeg$ peaked at $b\approx 9\arcdeg$.  Between $v=-15$~\kms\ and
$v=-9$~\kms\ the size of the shell is roughly constant.  At the central
velocity of $v=-12$~\kms\ the shell is brightest on the high longitude end
of the images and traces a faint arc through the low-longitude end.  Towards
larger velocities (i.e. towards the local gas) it decreases in angular size
and finally disappears at $v\approx-4$~\kms.  The caps of the shell, on both
the near and far sides are not readily detected.  On the near side, at
$v\sim-4$~\kms\ emission from the shell is obscured by absorption in the
Coalsack nebula and emission from the local gas.  The geometry of this shell
is best described as conical.  If there is emission from the far cap at $v>
30$~\kms, it is swamped by the \HI\ emission from the Sagittarius-Carina
spiral arm.  As seen in Figs.~\ref{fig:shell} \& \ref{fig:annotated}, the
emission void above the plane contains a vertical filament at $l\approx
306\arcdeg$, extending from $b\approx 3\fdg5$ to the top of the shell.  This
filament appears to correspond to the Galactic worm GW 305.5+9.4
\citep{koo92}.

Using the Galactic rotation model of \citet{fich89} and adopting the IAU
standard values for the Sun's orbital velocity, $\Theta_0 = 220$~\kms, and
Galactic center distance, $R_0 = 8.5$~kpc, we estimate the distance to GSH
304-00-12 from the central velocity of $v=-12$~\kms.  Assuming a velocity
error of $\pm 10$~\kms\ to account for streaming motions \citep{burton88},
we determine a distance of $D=1.2\pm1.0$ kpc.  At this distance the shell
has physical diameters of 280~pc and 200~pc in the longitude and latitude
directions, respectively.  The near edge of the Sagittarius-Carina arm is at
$1.26\pm0.23$ kpc \citep{seidensticker89}.  Although the shell distance
errors are large, the coincidence of the shell and spiral arm distances
suggests that the shell is at the edge of the arm.

GSH 304-00-12 has a velocity full width of $\Delta v = 23$~\kms.
Disentangling the spatial extent of the shell from the physical expansion is
difficult.  Because of Galactic rotation, structures that span hundreds of
parsecs will be observed over a range of radial velocities, even if they are
not expanding.  The velocity-distance relationship can be partially
separated by considering the velocity gradient along the line of sight.  At
the location of shell the velocity gradient is $\sim 11~{\rm
km~s^{-1}~kpc^{-1}}$.  Therefore, if the shell is roughly spherical, such
that the shell diameter along the line of sight is on the order of 240 pc,
it should extend over approximately 3~\kms\ in velocity space.  Given the
errors involved in deriving distances from the rotation curve, we believe
that the velocity expanse due to the shell's physical extent in negligible.
Therefore, we estimate the shell's expansion velocity as half of the
velocity full width, $v_{\rm exp} \approx \Delta v /2 = 12$~\kms.  This
expansion velocity is comparable to shells of similar size in the SMC and
somewhat smaller than shells in the LMC \citep{kimphd,stanimirovicphd}.

\begin{figure}
\begin{center}
\includegraphics[scale=0.65,angle=-90]{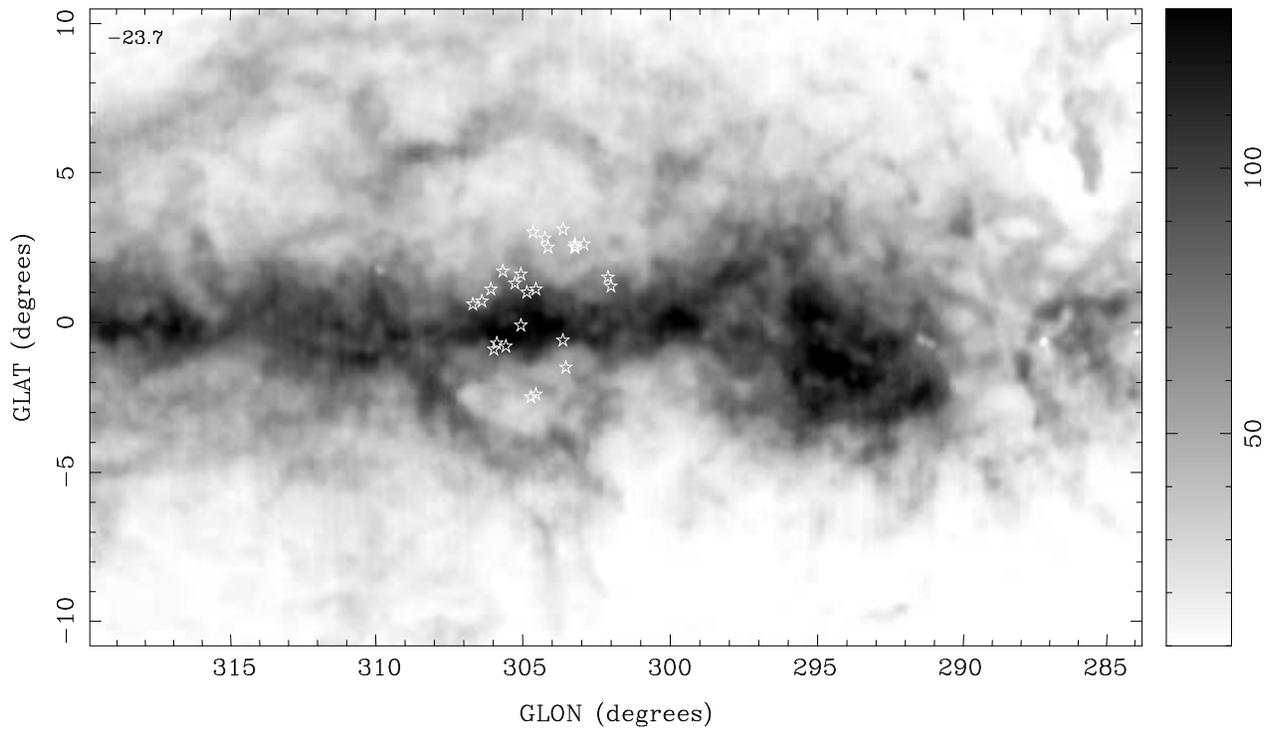} 
\figcaption[]{\HI\ channel image at $v=-24$~\kms\ of GSH~305+01-24 in the
edge of the Sagittarius-Carina arm.  The grey scale is linear from 10 to
130~K, as shown in the wedge at the right.  The shell, of angular diameter
$7\arcdeg \times 11\fdg3$, is centered at $l=305\fdg1$, $b=+1\fdg0$.  The
stars mark the positions of stars in the Cen OB1 association.
\citep{humphreys78}.
\label{fig:stars}}
\end{center}
\end{figure}

The second shell, GSH~305+01-24, has a smaller angular size of $7\arcdeg
\times 11\fdg3$ and is located at $l=305\fdg1$, $b=+1\arcdeg$, $v=-24$~\kms.
This shell is centered slightly above the plane and expands further above
and below the plane than it does along the plane.  An \HI\ image of
GSH~305+01-24 at $v=-24$~\kms\ is shown in Fig.~\ref{fig:stars} overlaid
with the positions of stars in the Cen OB1 stellar association.  The
relationship of the shell to the stellar association is discussed below.

Like the first shell, GSH~305+01-24 does not have a clearly defined near
cap.  It does, however, seem to have a far cap.  This shell appears to open
conically from within the spiral arm at $v\sim -34$~\kms.  It increases in
angular size towards larger velocities in the interarm region and reaches
its maximum size at $v=-24$~\kms.  The angular size is relatively constant
until $v\approx -20$~\kms, where the shell begins to close.  Beyond $v\approx
2-0$~\kms, emission from the shell is confused with that of GSH 304-00-12.
The central velocity of $v=-24$~\kms\ implies a distance of
$D=2.2\pm0.9$~kpc.  At this distance the shell has physical radii of $140
\times 220$~pc ($l \times b$).  GSH~305+01-24 has a velocity full width of
$\Delta v=14$~\kms.  Again, the velocity width due to the physical extent of
this shell along the line of sight is probably negligible so we calculate an
expansion velocity of $v_{\rm exp}\sim 7$~\kms.

\subsection{System Geometry}
\label{subsec:geometry}
The ISM along this line of sight is complicated.  Spectro-photometric
mapping of the region show that the Coalsack nebula is at an average
distance of $199\pm40$ pc and consists of at least two overlapping clouds at
188 and 243 pc \citep{seidensticker89}.  CO emission from the Coalsack
nebula is geometrically centered at $l=303\arcdeg$, $b=0\arcdeg$, covers
$\sim 30~{\rm deg}^2$ on the sky and extends at least over the systemic
velocity range $v_{rad} = -10$ to $+8$~\kms\ \citep{bronfman89}.  \HI\
self-absorption from gas in the Coalsack was detected by \citet{bowers80}
and is also observed in the SGPS dataset. The interarm region between the
Coalsack and the edge of the Sgr-Car arm is nearly devoid of dust and
molecular gas and the near edge of the spiral arm is at about 1.3 kpc
\citep{seidensticker89}.  \HI\ and CO emission between $l=290\arcdeg -
320\arcdeg$ and covering LSR velocities between $\sim -40$ and
$\sim-20$~\kms\ corresponds to the Sgr-Car arm at a mean distance of $\sim
2$~kpc.  For velocities in the range $v\approx-30$~\kms\ to
$v\approx-18$~\kms\ the spiral arm is nearly perpendicular to the line of
sight.  There is a very slight velocity gradient, such that the high
longitude end of the arm is at more negative velocities than the low
longitude end.

For a small range of velocities ($-27 \leq v \leq -20$~\kms) GSH~305+01-24
overlaps with GSH~304-00-12.  Though the shells are coincident in velocity
space, there is no reason to believe that they are physically overlapping.
The central velocities place the shells more than a kiloparsec apart and the
line of sight widths are insufficient to overlap the shells.  Also, the
morphology of the shells do not appear altered by an interaction.  The
overlap is most likely due to the expansion of the shells and not a true
interaction.

Both shells appear to have formed in the edge of the Sagittarius-Carina arm.
GSH~305+01-24 seems to close near the edge of the spiral arm, whereas
GSH~304-00-12 expands into the interarm region.  We suggest that
GSH~304-00-12 experienced exaggerated expansion into the interarm region due
to the pressure gradient leading away edge of the spiral arm.  This explains
the relatively constant size over the range of velocities covering the
interarm region and the lack of a detectable near cap.  GSH~305-00-12 also
extends into the range of velocities covered by the Coalsack nebula.  The
nebula at a distance of $199\pm40$ pc \citep{seidensticker89}, however, is
considerably closer than GSH~305-00-12.  Unless the distance to
GSH~304-00-12 has been significantly over-estimated or the shell's diameter
along the line of sight far exceeds its diameter in the plane of the sky, it
is unlikely that the shell is physically interacting with the Coalsack.  The
lack of emission from a near cap, though, could alternately be explained by
absorption by cold \HI\ in the Coalsack.

\subsection{Related Structures}
\label{subsec:related_structures}
The lower portion of GSH~305+01-24 was previously identified as an \HI\
shell associated with the Wolf-Rayet star $\theta$ Muscae \citep{cappa84}.
However, given the larger field of view in the current observations, this
shell arc appears connected with the arc above the plane.  The void of
GSH~305+01-24 is bisected by emission from the Galactic plane.  Because the
rotation curve is dual-valued at these velocities, some emission is likely
due to much more distant gas located on the far side of the tangent point.
In the context of these measurements the $\theta$ Muscae shell must be part
of the larger structure that we identify as GSH~305+01-24.  \citet{cappa84}
also noted a second \HI\ depression at $v=-12$~\kms\ that we identify as
part of GSH~304-00-12.

\begin{figure}
\begin{center}
\includegraphics[scale=0.7,angle=-90]{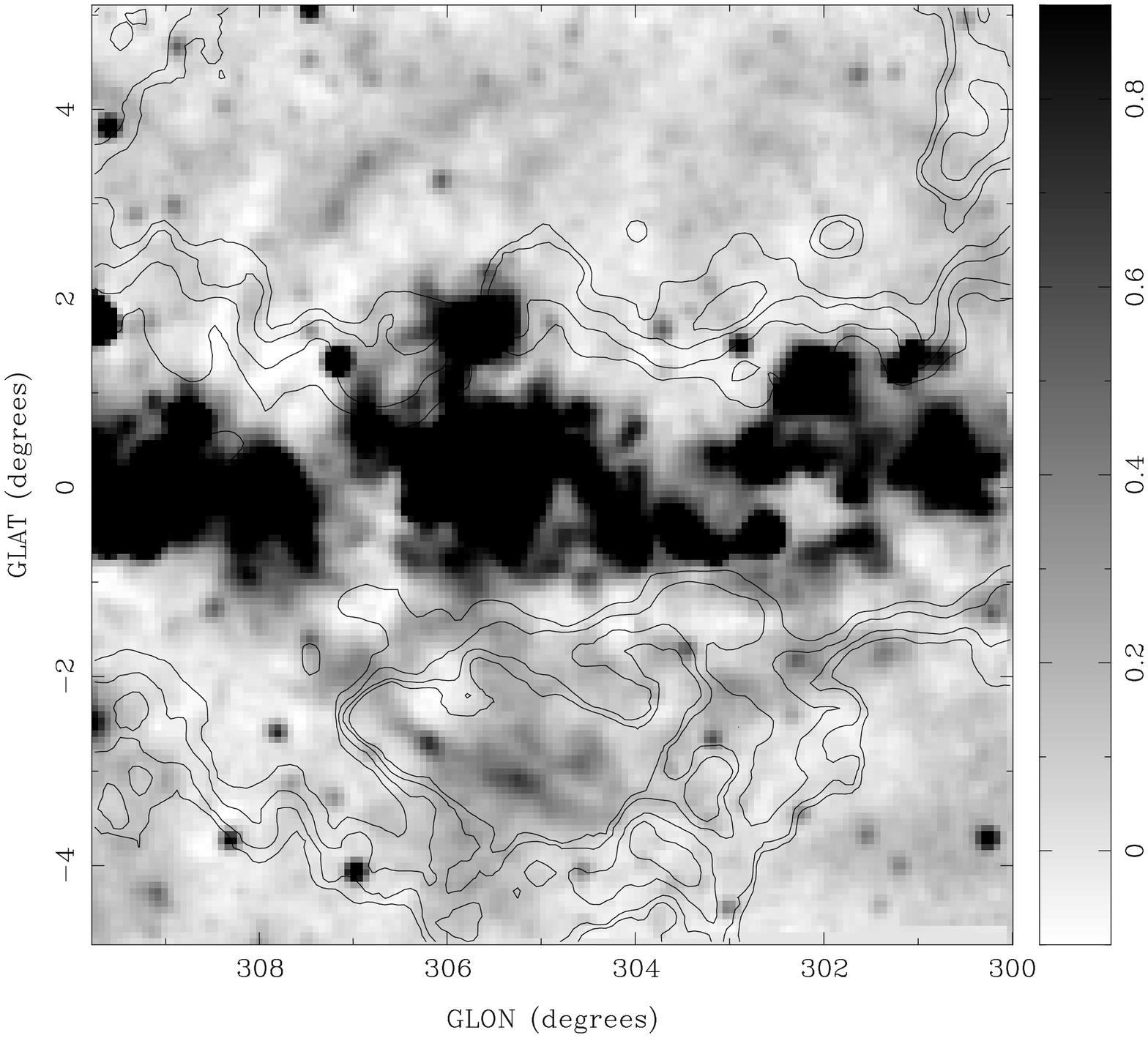}
\figcaption[]{Grey scale image of the 2.4 GHz continuum emission of the
GSH~305+01-24 region overlaid with \HI\ contours at $v=-24$~\kms.  The grey
scale is linear and goes from $0.1$ to $0.9~{\rm mJy~Bm^{-1}}$ as shown in
the wedge at the right.  The continuum image is from the \citet{duncan95a}
survey of the Galactic Plane and has been filtered to enhance small scale
structure as described therein.  The \HI\ contours are at 45, 50, 60, and 70
K of $T_b$.  The faint arcs of continuum emission in the lower half of
GSH~305+01-24 trace the interior edge \HI\ shown as shown by the contours.
\label{fig:coalsack2.4}}
\end{center}
\end{figure}

There is 2.4~GHz continuum emission corresponding to the lower half of
GSH~305+01-24.  Fig.~\ref{fig:coalsack2.4} is a 2.4 GHz continuum image from
the \citet{duncan95a} survey of the Southern Galactic Plane overlaid with
contours of an \HI\ channel image of the shell at $v=-24$~\kms.  The
continuum emission excellently traces the inside edge of the shell below the
plane.  Also, the continuum may trace a portion of the shell above the plane
near $l=301\arcdeg$, though this emission is very weak making a definitive
detection difficult.  There is an arc of continuum emission through the the
upper \HI\ void that does not correspond to any \HI\ features.  These arcs
were identified by \citet{duncan95a} as a possible supernova remnant,
G303.5+0.  However, close examination of IRAS\footnote{See IRAS Sky Survey
Atlas Explanatory Supplement at
http://www.ipac.caltech.edu/ipac/iras/issa.html} 100 \micron\ images of the
region reveals a very faint infrared shell coincident with the radio
continuum features.  These features were also identified as an H$\alpha$
shell, the Coalsack Loop, by \citet{walker98}.  The detection of the arcs in
the infrared implies that the emission is thermal and therefore suggests
that the arcs are not part of a supernova remnant.  The close morphological
similarity of the lower arcs and the \HI\ shell leaves little doubt that the
two are associated.  We point out that the Coalsack Loop is much larger in
angular size than the Coalsack nebula so that even though the Loop is behind
the nebula, extinction in the nebula will not affect the H$\alpha$ emission.
We believe, therefore, that the continuum emission is the photo-ionized
region interior to the \HI\ shell.

\subsection{Is GSH~305+01-24 Associated with Centaurus OB1?}
\label{subsec:cenob1}
The Cen OB1 stellar association is centered at $l=304\arcdeg$, $b=+0\fdg5$
with radial velocities in the range $-38$~\kms\ to $-10$~\kms.  The adopted
distance is 2.5~kpc \citep{humphreys78}, which puts the cluster, to within
the errors, at the same distance as GSH~305+01-24.  The positions of the
stars are plotted on a channel image of GSH~305+01-24 at $v=-22$~\kms\ in
Fig.~\ref{fig:stars}.  The morphology of the shell appears to trace the
stellar distribution, suggesting an association.  The shell extends further
above the plane than below, a characteristic that is matched by the stellar
distribution.  The cluster contains 21 stars with measured luminosities and
temperatures and calculated ages and masses \citep{humphreys78,kaltcheva94}.
The earliest stars in the association are of type O9, which suggests that
more massive stars may have already exploded as supernovae.  A Wolf-Rayet
star, $\theta$ Muscae, of spectral type WC6 is also in the cluster.  This
system is a binary, possibly a triple, consisting of the W-R+O close binary
and a distant O supergiant \citep{hartkopf99}.  A priori, it seems likely
that this association is the energy source for the GSH~305+01-24.  

To explore whether GSH~305+01-24 could be explained as a wind-driven bubble
associated with Cen OB1, we employed a very simple one-dimensional numerical
model to study the evolution of a shell powered by the stars in this
association.  The model follows standard equations for the evolution of a
pressure-driven wind bubble \citep[e.g.][]{weaver77,oey95}.  The system is
simplified to contain a centrally located power source, in this case the OB
association, that supplies a time-dependent wind luminosity which drives the
expansion of a shell.  We assume that the internal pressure and temperature
are constant throughout the bubble interior.  The system is further
simplified by assuming an ideal, monatomic gas equation of state, and a
cool, homogeneous external medium.  The bubble expansion can be described by
three first-order equations: the rate of expansion of the shell, the
equation of momentum conservation and the equation of energy conservation.
The rate of expansion, $v_{\rm exp}$, is simply the first-derivative of the
shell radius, $R_{sh}$,
\begin{equation}
\frac{dR_{sh}}{dt} = v_{\rm exp},
\label{eq:radius}
\end{equation}
and the momentum conservation equation, including the effects of external
and internal pressure, $P_e$ and $P_i$, respectively, is
\begin{equation}
\frac{dv_{sh}}{dt} = \frac{4\pi R_{sh}^2}{M_s}~\left( P_i - P_e -\rho_e
v_{\rm exp}^2\right),
\label{eq:velocity}
\end{equation}
where $M_s$ is the mass of the shell and $\rho_e$ is the density of the
ambient medium.  The equation of energy conservation is
\begin{equation}
\frac{dE_{sh}}{dt} = L_w - P_i \frac{dV}{dt},
\label{eq:energy}
\end{equation}
where $L_w$ is the wind luminosity and $V$ is the bubble volume,
$V=(4\pi/3)R_{sh}^3$.  The internal energy of the shell, $E_{sh}$ is related to
the internal pressure according to the relation
\begin{equation}
E_{sh} = \frac{P_i V}{\gamma - 1},
\label{eq:pressure}
\end{equation}
where the adiabatic index is $\gamma =5/3$.  The wind luminosity, $L_w = 1/2
\dot{M} v_{\infty}^2$, is calculated from an estimate of the mass-loss rate,
$\dot{M}$, and escape velocity, $v_{\infty}$, of each star in the Cen OB1
association.  The external pressure is given by $P_e = \rho_e
c_s^2\gamma^{-1}$, where $c_s$ is the local speed of sound in the external
medium, which is assumed to be $\sim 10$~\kms\ for a warm neutral medium at
$T\sim 5000$ K.  Using temperatures, luminosities, masses, and ages for 21
Cen OB1 stars tabulated by \citet{kaltcheva94}, we estimate the mass-loss
rate from the empirically determined mass-loss rates of OB stars made by
\citet{jager88}.  The terminal velocity is estimated from the OB stellar
wind model of \citet{leitherer92}.  Using the ages of the stars we calculate
the total wind luminosity as the sum of all stars that exist at each time
step.  Only stars for which masses, luminosities, and ages are available are
included in the model.  The spectral type, escape velocities, mass-loss
rates, and wind luminosities are given in Table~\ref{tab:stars}.  For
simplicity, we assume a constant external density, $\rho_e = 1.67\times
10^{-24}~{\rm gm~cm^{-3}}$, corresponding to one hydrogen atom per cubic
centimeter.  The model is calculated with time steps of $1.5\times 10^3$~yr
and the initial conditions are calculated from the self-similar solutions to
the differential equations at the first time step, $t=1.5\times 10^3$~yr.

\begin{figure}
\begin{center}
\includegraphics[scale=0.7]{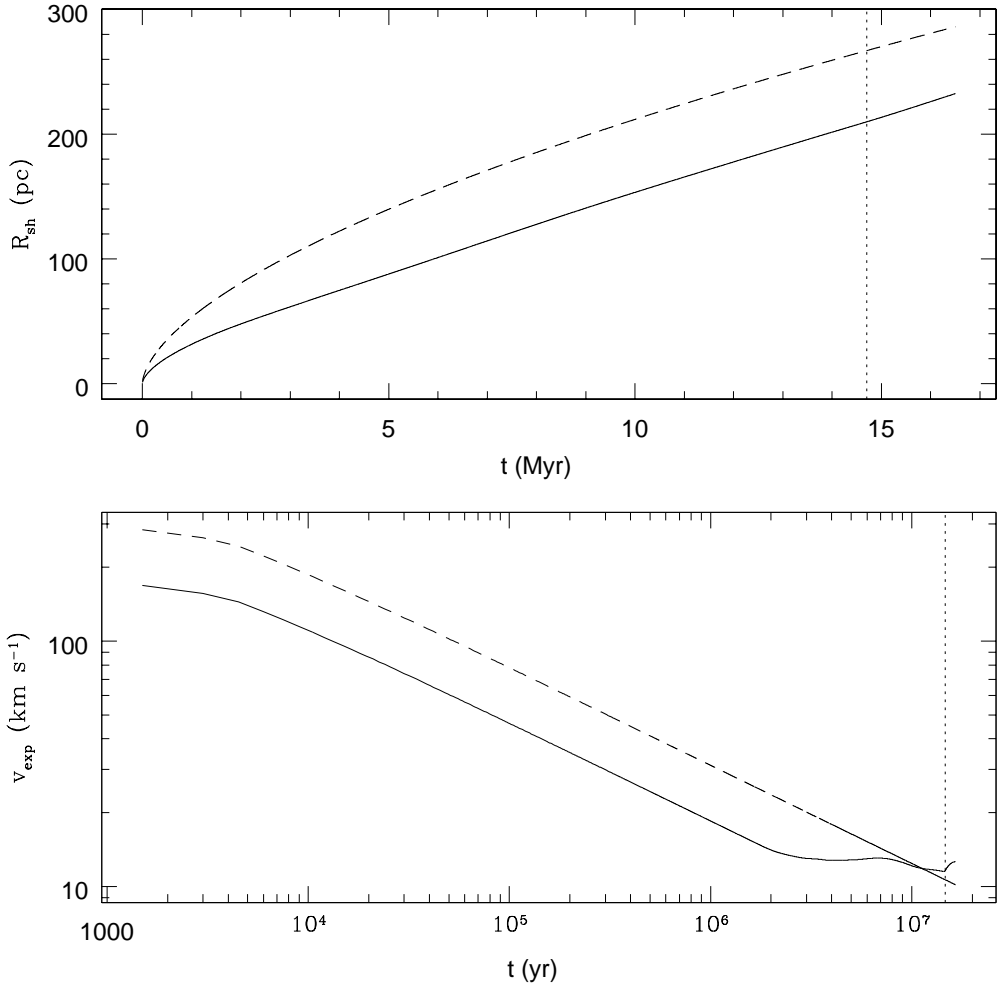}
  \figcaption[]{Simple numerical model of the shell radius and velocity
  evolution calculated for a constant luminosity input equal to the sum of
  the Cen OB1 stellar wind luminosities (dashed line) and for a time
  dependent total wind luminosity calculated from the ages, masses and
  luminosities of the Cen OB1 population (solid line).  The vertical dotted
  line marks the present age of the stellar association.  The actual shell
  radius and expansion velocity are well modeled by a time-dependent wind
  luminosity.  The flattening of the velocity curve near 14 Myr in the time
  dependent model is due to the Wolf-Rayet star. 
\label{fig:model}}
\end{center}
\end{figure}

The results of the numerical integration are shown in
Fig.~\ref{fig:model}. We explored two cases: one where the luminosity
increases with time as stars are ``born'' in the association and another
case of constant luminosity equal to the sum of all stars, regardless of
age.  The time-dependent wind luminosity case is plotted with a solid line
and the constant wind luminosity case is plotted with a dashed line.  The
vertical dotted line marks the age of the oldest star in the association
(14.7 Myr), with time $t=0$ assumed to be the birth of that star.  The
expansion velocity flattens near 14 Myr because of the high mass-loss rate
of the Wolf-Rayet star, $\theta$ Muscae.  Several differences between the
two models are of note.  First, the shell radius for the constant wind
luminosity case can be fit as $R_{\rm sh} \propto t^{3/5}$, the analytic
self-similar solution to the problem.  But, in the time-dependent wind
luminosity case the radius goes as $t^{3/5}$ initially and then is better
fit as $R_{sh}\propto t$.  Second, the usual assumption of coeval star
formation for stellar associations leads to a slight over-estimate of the
total energy input in a wind blown bubble.  This effect is most noticeable
as an overall increase in final shell radius by a factor of $\sim 1.4$;
expansion velocity is relatively insensitive to the total energy input for
larger ages.  It seems, then, that including the stellar ages in the wind
luminosity model results a marginally better fit to the data.

We find that GSH~305+01-24 is surprisingly well described by the
\citet{weaver77} model of a wind-blown bubble with a time-dependent wind
luminosity from Cen OB1 stars.  This model predicts a shell radius of $\sim
210$~pc, which corresponds well with the measured radii of $140\times
220$~pc; whereas the time-independent luminosity model predicts a radius of
$\sim 300$~pc.  The model predicts an expansion velocity of $\sim 11$~\kms\
for both the time independent and dependent luminosity cases, which also
agrees well with an inferred expansion velocity of $\sim 6$~\kms.  The good
agreement suggests that GSH~305+01-24 was indeed formed from the stellar
winds of the Cen OB1 association.  The size and expansion velocity are
consistent with a shell age approximately equal to the age of the oldest
star in the association ($\sim 15$~Myr).  \citet{kaltcheva94}, however,
derive an age of $8.9\pm3.1$~Myr for Cen OB1, from the average of the
population's ages.  The stellar ages are evenly distributed from 14.7 Myr to
3.6 Myr with no evidence for sequential star formation, but rather there is
evidence for continuous star formation, which makes it difficult to define
an age for the cluster.  Using the wind luminosities and ages of the Cen OB1
association stars we find the total energy released in the form of stellar
wind luminosity during the lifetime of the association is $\sim 8.3\times
10^{51}$ ergs.  We can compare this to the expansion energy of the shell -
the equivalent energy instantaneously deposited at the center of the shell -
derived by \citet{chevalier74} and given in equation 2 of \citet{heiles79}:
$E_E = 5.3\times 10^{43} n_0^{1.12} v_{\rm exp}^{1.4} R_{sh}^{3.12}$.  To
account for the current radius and rate of expansion of GSH~305+01-24, the
expansion energy is $\sim 9.8 \times 10^{51}$ ergs, which agrees well with
the total energy of the Cen OB1 stellar winds.

\subsubsection{Limitations of the Model}
The model shows, as it was intended to do, that the origin of GSH 305+01-24
can be explained, with respect to the energetics, as stellar wind luminosity
from the Cen OB1 association.  Despite the good reproduction of the shell
characteristics by this model, it should be stressed that it is a
over-simplification and cannot replace full magnetohydrodynamical
simulations of \HI\ shells from OB associations.  No attempts have been made
to include multi-dimensional modeling, radiative cooling, magnetic fields,
or the spatial distribution of the energy input sources.  Additionally, the
model contains no morphological information that can be helpful when trying
to fully understand shell dynamics.  Here we summarize some of the
components of a non-idealized ISM that have not been included in the
calculations and how those might effect the results.

The model neglects energy injection from supernovae because there is no
direct evidence for any associated with this cluster.  However, it is
possible that there have been supernovae in the association.  The effect of
supernovae on the wind-driven bubble evolution is very similar to that of
stellar winds as an input energy source \citep{mccray87,chu90}.
\citet{mccray87} show in their equation 1 that for a standard initial mass
function, an association of $N_{\star}$ stars with mass greater than $7~{\rm
M_{\odot}}$, supernovae input a mean luminosity over the association
lifetime of $P \approx 6.3 \times 10^{35}~{\rm erg s^{-1}} (N_{\star}
E_{51})$.  However, only $\sim 20$\% is available in the first 10 Myr.
Therefore, if $N_{\star} \sim 15$, supernovae would have increased the
modeled radius and expansion velocity of GSH 305+01-24 by about a factor of
two.

As shown in Fig.~\ref{fig:stars} the stars of the Cen OB1 association are
distributed throughout the interior of GSH~305+01-24.  The model, however,
assumes all stars are at the center of the shell.  The B stars may have
migrated as much as $\sim 50$ pc in the shell life-time \citep{mccray87},
which suggests that the stars could have been initially distributed over a
region $\lesssim 100$ pc in diameter.  This inferred remaining distribution
should be negligible on the calculated shell radius and expansion velocity,
but could affect the morphology of the shell.

The shell morphology would likely be dominated by external density
variations that were not included in this model.  As shown in studies of
supershell expansion in stratified media \citep[e.g.][]{maclow89}, shells
expanding out of the plane of the Galaxy experience exaggerated expansion
and blow-outs in that direction.  For GSH~305+01-24, the shell radius is on
the order of the scale height of the \HI\ disk, at which height the density
gradient of the disk becomes significant.  Also, as noted above, the shells
are at the edge of the Sagittarius-Carina spiral arm, so the density and
pressure should decrease into the interarm region.  Both of these effects
will undoubtedly lead to exaggerated expansion down both density gradients.
Small scale density fluctuations and the development of instabilities in the
course of shell expansion can also effect the shell shape and size and can
lead to break-outs in localized regions.  The effect of such break-outs
would be to equalize the internal and external pressures and stall
expansion.

Finally, three dimensional simulations of shell expansion in multi-phased,
magnetized, turbulent media show shells that are far from circular and
exhibit many merging cavities and tunnels rather than a fully empty void
\citep{korpi99}.  Uniform magnetic fields in the disk can act to temporarily
confine shells to the plane, and thereby affect the shell radius
\citep{tomisaka98}.  Clearly, the model employed here has numerous
limitations, but it adequately shows that winds from the Cen OB 1
association can account for the observed shell radius and expansion
velocity.

\section{Conclusions}
\label{sec:concl}
We have shown images of two new \HI\ shells, GSH~305+01-24 and
GSH~304+00-12, located at the edge of the Sagittarius-Carina spiral arm in
the direction of the Coalsack nebula.  Both shells are very large, with
radii on the order of 200~pc.  We employed a simple time dependent wind
luminosity model based on the ages and luminosities of the stars in the Cen
OB1 association to describe the evolution of GSH~305+01-24.  The model
predictions agree well with the observed shell radius and expansion
velocity.  We use this agreement, the positional coincidence of the shell
and OB association, and the similarities between the shell morphology and
the stellar distribution to argue that GSH~305+01-24 is a wind-blown bubble.
This shell is associated with a 2.4~GHz radio, infrared and H$\alpha$ shell
that were previously posited to be an old supernova remnant
\citep{duncan95a,walker98}.  Because of the infrared detection of the shell
we believe that the emission is thermal and therefore not part of a
supernova remnant, but instead emission from the shell itself.

We also found a large angular diameter shell, GSH~304-00-12, that does not
appear to have formed from any known stellar associations.  This shell is
embedded in the edge of the Sagittarius-Carina arm and expanded into the
interarm region, resulting in a conical geometry.  We find no corresponding
emission features in other wavebands.  GSH~304-00-12 has no detectable near
cap, but the velocities where the cap is expected overlap velocities in the
Coalsack nebula.  We believe that there is no physical association between
the nebula and the shell, but we suggest emission from the cap may be
absorbed by cold \HI\ in the nebula.  Additionally, although GSH~305+01-24
and GSH~304-00-12 overlap in velocity space, there is no direct evidence for
any interaction between them.
\acknowledgements N.\ M.\ M.-G.\ would like to thank E.\ J.\ Hallman for
helpful conversations and code.  We would also like to thank the anonymous
referee for helpful suggestions that improved the paper.  J.\ M.\ D.\ and
N.\ M.\ M.-G.\ acknowledge support of NSF grant AST-9732695 to the
University of Minnesota.  N.\ M.\ M.-G.\ is supported by NASA Graduate
Student Researchers Program (GSRP) Fellowship NGT 5-50250.  B.\ M.\ G.\
acknowledges the support of NASA through Hubble Fellowship grant
HST-HF-01107.01-A awarded by STScI, which is operated by AURA Inc. for NASA
under contract NAS 5-26555.

\small 

\normalsize

\clearpage
\begin{deluxetable}{lcccccc}
\tabletypesize{\scriptsize}
\tablewidth{0pt}
\tablecaption{Properties of massive stars in the Cen OB1
stellar association included in the numerical model for GSH 305+01-24.
\label{tab:stars}}
\tablehead{
\colhead{Star} & \colhead{Spectral Type\tablenotemark{a}} & \colhead{Age} &
\colhead{$v_{\infty}$} &  \colhead{Log $\dot{M}$} & \colhead{$L_w$} & \colhead{$E_{tot}$\tablenotemark{b}}\\
\colhead{} & \colhead{} & \colhead{(Myr)} & \colhead{(\kms)} &
\colhead{(${\rm M_{\sun}~yr^{-1}}$)} &
\colhead{($10^{35}~{\rm erg~s^{-1}}$)} & \colhead{($10^{49}~{\rm erg}$)} 
}
\startdata
HD 110639  & B1 Ib   & 10.3   &   1514.1   &  -6.22 &  4.35    &    14.15  \\ 
HD 111613  & A2 Iab  & 11.9   &   893.8   &  -6.10 &  2.00     &   7.51 \\  
HD 111904  & B9 Ia   & 9.6    &   979.5   &  -5.30 &  15.15   &     45.91 \\  
HD 111934  & B2 Ib   & 9.1    &   1427.3   &  -5.60 &  16.13     &   46.31 \\  
HD 111973  & B5 Ia   & 8.6  &     1169.3   &  -5.91 &  5.30     &   14.39 \\   
HD 111990  & B3 Ib   & 11.0   &   1341.4   &  -5.40 &  22.58    &    78.37 \\   
HD 112272  & B0.5 Ia  & 4.7   &   1563.9   &  -5.52 &  23.28     &   34.52 \\   
HD 112366  & B9 Ia-Ia & 10.0   &  1044.1   &  -5.40 &  13.68    &    43.16 \\  
HD 112364  & B0.5 Ib  & 12.0   &  1607.1   &  -5.90 &  10.25     &   38.81 \\   
HD 112842  & B3 Ib   & 14.7  &    1201.5   &  -5.40 &  18.11   &     84.02 \\   
HD 113012  & B0 Ib   & 12.6  &    1764.2   &  -6.22 &  5.91     &   23.50 \\   
HD 113422  & B1 Ia  & 6.5     &   1423.9   &  -5.92 &  7.68    &    15.75  \\  
HD 114011  & B1 Ib  & 12.9     &  1637.7   &  -6.14 &  6.12     &   24.92  \\  
HD 114213  & B1 Ib  & 9.5     &   1320.5   &  -6.22 &  3.31    &    9.93 \\   
HD 115704  & B0.5 Ia-Iab & 3.7  & 1568.0   &  -6.15 &  5.49    &    6.40 \\   
HD 116119  & A0 Ia  & 10.2    &   1006.3   &  -7.16 &  0.22     &   0.71 \\   
HD 115363  & B1 Ia  & 3.6     &   1494.8   &  -5.27 &  37.82    &    42.96 \\   
HD 114122  & B0.5 Ia-Iab & 7.6  & 1560.5   &  -6.15 &  5.43    &    13.03 \\    
HD 114886  & O9 II-III & 4.9   &  2161.6   &  -5.57 &  39.64     &   61.29 \\    
HD 113421  & B0.5 III  & 6.7   &  2386.8   &  -6.90 &  2.26     &   4.78 \\  
HD 113708  & B0.5 III  & 6.2  &   2197.7   &  -6.90 &  1.92    &    3.75 \\   
WR 48\tablenotemark{c}   &  WC6+O9.5I &  \nodata &  2000.0 & -3.74  &   2294.0    &    217.17 \\   
\enddata 
\tablenotetext{a}{From \citet{humphreys78}.}
\tablenotetext{b}{Total lifetime energy input, $E_{tot} = L_w t$, where
$t= {\rm age}$}
\tablenotetext{c}{The W-R+O9.5 close binary is assumed to exist in its
current state of high mass loss for $\sim 3\times 
10^5$~yr.  The mass loss rate and
terminal velocity are from \citet{leitherer97}.}
\tablecomments{Only stars with observed temperatures, luminosities, masses and
ages were used.  Terminal velocities were estimated from the OB stellar wind
model of \citet{leitherer92}.  The mass-loss rates are based on the
empirically determined mass-loss rates of OB stars in \citet{jager88}.  In
the model, stars are ``turned on'' at 14.7 Myr minus their age. }
\end{deluxetable}

\end{document}